\documentstyle[preprint,prd,aps]{revtex}
\begin{document}
\draft

\title {Symplectic Gravity Models in\\
Four, Three and Two Dimensions}

\author{Oleg Kechkin and Maria Yurova}

\address{Nuclear Physics Institute,\\
Moscow State University, \\
Moscow 119899, RUSSIA, \\
e-mail: kechkin@cdfe.npi.msu.su}

\date{October 1996}

\maketitle

\draft
\begin{abstract}
A class of the $D=4$ gravity models describing a coupled system of $n$
Abelian vector fields and the symmetric $n \times n$ matrix generalizations
of the dilaton and Kalb--Ramond fields is considered. It is shown that the
Pecci--Quinn axion matrix can be entered and the resulting equations of
motion possess the $Sp(2n, R)$ symmetry in four dimensions.
The stationary case is studied. It is established that the theory allows a
$\sigma$--model representation with a target space which is invariant under
the $Sp[2(n+1), R]$ group of isometry transformations. The chiral matrix of
the coset \mbox{$Sp[2(n+1), R]/U(n+1)$} is constructed. A K\"ahler formalism
based on the use of the Ernst \mbox{$(n+1) \times (n+1)$} complex symmetric
matrix is developed.
The stationary axisymmetric case is considered. The Belinsky--Zakharov
chiral matrix depending on the original field variables is obtained. The
Kramer--Neugebauer transformation, which algebraically maps the original
variables into the target space ones, is presented.
\end{abstract}
\pacs{PACS numbers: 04.20.Jb, 04.50.+h}

\draft

\narrowtext
\section{Introduction}
Last years much attention had been attracted by the study of symmetries
for gravity models
arising as the low energy limit of superstring theory \cite {mahs}--\cite
{bkr}. This activity was motivated by the problems of supersting theory,
as well as by the remarkable results obtained for the
Einstein and Einstein--Maxwell theories in three and two dimensions. Namely,
it had been established the chiral matrix formulations for the latter
theories in the stationary case, and their complete
integrability in the stationary axisymmetric case \cite {g1}--\cite {c}.

Recently it was shown that the simplest string gravity model,
the Einstein--Maxwell--Dilaton--Axion (EMDA) system,
can be represented as the $2 \times 2$ matrix generalization of
the pure Einstein theory \cite {gk1}--\cite {ky}.
It means that one can exchange the complex symmetric $2 \times 2$ matrix,
describing the EMDA theory in three dimensions, by the Ernst potential of
the Einstein theory to obtain this last one from the former system
\cite {gk1}. Thus, the mentioned matrix (the Ernst matrix potential)
generalizes the Ernst potential to the EMDA case, and both the Einstein
and EMDA theories belong to the same class of gravity models. In \cite {ky}
it was shown that the established analogy is even the more complete one,
and can be naturally prolonged to the stationary axisymmetric case. It
had been shown, that the EMDA system, as the Einstein theory, allows both
the target space representation and the formulation based on the use of the
original field components. These two formulations are differentially related
in three dimensions and can be algebraically mapped one into another in two
dimensions \cite {ky}. This map is known as the Kramer--Neugebauer
transformation for the stationary axisymmetric Einstein equations \cite {kn};
its existence is an important property of the theory \cite {bm}.

The formal properties of the EMDA theory do not actually depend on the Ernst
matrix potential dimension and it is possible to consider a series of
similar
gravity models in three and two dimensions. All these models exist as
the chiral ones and possess the remarkable features of
the Einstein theory. But first of all one must give the answer
to the question what kind of the four--dimensional theories being
reduced to three and two dimensions lead to such models.

In this paper we present such a class of four--dimensional theories. Any
theory generalizes the EMDA system by the following way: here the dilaton
and Kalb--Ramond fields become the $n \times n$ symmetric matrices, and
instead of the Maxwell field one has the column of $n$ Maxwell fields
(thus, the EMDA theory corresponds to the case of $n = 1$).

It is shown that one can introduce the Pecci--Quinn axion
matrix for an arbitrary $n \ge 1$ and that the resulting equations of motion
possess the $Sp(2n, R)$ hidden symmetry. This fact generalizes the
one for the EMDA theory which $Sp(2, R) \sim SL(2, R)$ symmetry is known
as the $S$--duality \cite {stw}--\cite {ko}. Also it is shown that the
model admits the discrete symmetry transformation which had been named as
the ``strong--weak coupling duality'' for the low energy heterotic
string theory with moduli fields in four dimensions \cite {s4}.

In the next section we consider the stationary case and reduce the
four--dimensional model with an arbitrary $n$ to three dimensions. We use
an approach which was firstly formulated by Israel and Wilson for the
Einstein--Maxwell theory in \cite {iw}, and recently had been generalized
to the EMDA theory in \cite {ggk}. It is shown that the system under
consideration allows the $\sigma$--model representation with a symmetric
target space, which is a natural matrix generalization of the EMDA one.

Then, the K\"ahler formalism based on the use of the $(n + 1) \times (n
+ 1)$ Ernst matrix potential is developed. It is shown that the motion
equations become the matrix--valued Ernst equation. The chiral $2(n + 1)
\times 2(n + 1)$ matrix constructed on the real and imaginary parts of the
Ernst matrix potential is presented. This matrix belongs to the coset
$Sp[2(n + 1), R]/U(n + 1)$, and the established chiral formulation
possesses the $Sp[2(n + 1), R]$ group of symmetry transformations. The
corresponding transformations for the Ernst matrix potential have the
``matrix--valued $SL(2, R)$ form'' and directly generalize the
transformations for the Einstein theory. Also it is shown that the
$Sp(2n, R)$ symmetry, corresponding to the one of the four--dimensional
motion
equations, exists as the subgroup of the complete $Sp[2(n + 1), R]$ symmetry
group. At the end of the section the discrete transformation, which is the
analogy of the strong--weak coupling duality for the low energy
heterotic string theory in three dimensions \cite {s3}, is presented.

In the following section we demonstrate some new features of the model,
which arise in the stationary axisymmetric case. It is established that the
theory allows the alternative Ernst--like formulation using two real
$(n + 1) \times (n + 1)$ matrices, which are the nondualized
(i.e., depended on the original non--target space field components)
analogies of the Ernst matrix and its complex conjugation. These two
real matrices are used for the construction of the new
$2(n + 1) \times 2(n + 1)$ chiral matrix which coincides with the
Belinsky--Zakharov one for the case of the Einstein theory (see also \cite
{ky} for
the EMDA theory case). This new matrix undergoes the same $Sp[2(n + 1), R]$
transformations as the coset chiral matrix under the action of the isometry
group. The above mentioned real Ernst--like matrix potentials transform
as their complex Ernst analogies.

At the end of the paper we present the Kramer--Neugebauer map for an
arbitrary $n$. It turned out that this transformation has an especially
simple form  in terms of the real and complex pairs of the
Ernst matrix potentials.
\section{Four--Dimensional Equations}
In this paper we consider a class of gravity models with the action
\begin{equation}
^4S = \int d^4x {\mid g \mid}^{\frac {1}{2}} \{ - R+
Tr [ \frac {1}{2}(\partial p p^{-1})^2 - pFF^T + \frac {1}{3}(pH)^2] \},
\end{equation}
where $R$ is the Ricci scalar for the metric $g_{\mu \nu}$,
$(\mu = 0, ...,3)$; $F_{\mu \nu} = \partial _{\mu}A_{\nu} -
\partial _{\nu}A_{\mu}$ and
\begin{equation}
H_{\mu \nu \lambda} = \partial _{\mu}B_{\nu \lambda}
- \frac {1}{2} (A_{\mu}F^T_{\nu \lambda} + F_{\nu \lambda}A^T_{\mu})
+ {\rm cyclic}.
\end{equation}
Here the symmetric $n\times n$ matrix $p$ is constructed on the scalar
field components ($p = e^{-2\phi}$ for the EMDA case, where $\phi$ is the
dilaton field), $B_{\mu \nu}$ is the symmetric $n\times n$ matrix which
contains the antisymmetric tensor Kalb--Ramond fields (i.e., $B^T_{\mu \nu}
= B_{\mu \nu}$ and $B_{\mu \nu} = - B_{\nu \mu}$), and $A_{\mu}$ is the
$n\times 1$ column of the Abelian vector fields. This action describes the
Einstein and EMDA theories in the cases of $n = 0$ and $n = 1$ respectively
and provides their natural generalization for an arbitrary $n$. The models
of such type arise in the low energy limit of the heterotic string theory
after the compactification of extra dimensions on a torus
\cite {mahs}--\cite {bkr}.

As it has been done for the EMDA theory and for the string gravity with
moduli fields, one can establish an alternative formulation of the
problem which is connected with the dualization of the axion field
$H_{\mu \nu \lambda}$ in four dimensions. Namely, using
the first equation of the Euler--Lagrange ones,
\begin{equation}
\nabla ^{\mu}(pH^{\mu \nu \lambda}p)
= 0,
\end{equation}
\begin{equation}
\nabla _{\mu}(pF^{\mu \nu}) + \frac {1}{2}pH^{\mu \nu \lambda}F_{\mu \nu}
= 0,
\end{equation}
\begin{equation}
\nabla J^p + pFF^T - \frac {2}{3}(pH)^2
= 0,
\end{equation}
\begin{equation}
R_{\mu \nu} = Tr [\frac {1}{2}J^p_{\mu}J^p_{\nu} - 2p(F_{\mu \lambda}
{F^T}^{\nu \lambda} - \frac {1}{4}g_{\mu \nu}FF^T) +
pH_{\mu \lambda \sigma}pH_{\nu}^{\lambda \sigma} - \frac {1}{3}g_{\mu \nu}
(pH)^2]
\end{equation}
where $J^p = (\nabla p)p^{-1}$, it is possible to introduce the symmetric
pseudoscalar matrix $q$,
\begin{equation}
H^{\mu \nu \lambda} = \frac {1}{2}E^{\mu \nu \lambda \sigma}
p^{-1}q_{,\sigma}p^{-1}
\end{equation}
generalizing the Pecci--Quinn axion field.
It satisfies the relation
\begin{equation}
\nabla J^q - J^pJ^q + p\tilde FF^T = 0,
\end{equation}
where $J^q = (\nabla q)p^{-1}$ and $\tilde F^{\mu \nu} =
\frac {1}{2}E^{\mu \nu \lambda \sigma}F_{\lambda \sigma}$, which together
with Eqs. (4)--(6), rewritten using Eq. (7) as
\begin{equation}
\nabla _{\mu}(pF^{\mu \nu} + q\tilde F^{\mu \nu}) = 0,
\end{equation}
\begin{equation}
\nabla J^p + (J^q)^2 + pFF^T = 0,
\end{equation}
\begin{equation}
R_{\mu \nu} = Tr [\frac {1}{2}(J^p_{\mu}J^p_{\nu} +
J^q_{\mu}J^q_{\nu}) -2p(F_{\mu \lambda}{F^T}_{\nu}^{\lambda} -
\frac {1}{4}g_{\mu \nu}FF^T)]
\end{equation}
form the complete system for the action
\begin{equation}
^4S = \int d^4x {\mid g \mid}^{\frac {1}{2}} \{ - R +
Tr [ \frac {1}{2}((J^p)^2 + (J^q)^2) - pFF^T - q\tilde FF^T]\}.
\end{equation}
The action of this type firstly was considered by Breitenlohner, Gibbons
and Maison in \cite {bgm}, where some remarkable properties of the
model had been established.

The main purpose of this paper is to show a complete formal analogy
between the series of models describing by the action (1), and the
Einstein theory in three and two dimensions. However, the simplest
representative of the series, which possesses all the typical properties
in four dimensions, is the EMDA theory. Namely, let us establish the
symmetries of the theory for an arbitrary $n \ge 1$. To do it one can
introduce the complex variables
\begin{eqnarray}
z &=& q + ip,\\
{\cal F} &=& \frac {1}{2}(F + i\tilde F).
\end{eqnarray}
Then the set of motion equations obtains the form:
\begin{equation}
\nabla _{\mu}(z{\cal F}^{\mu \nu} - \bar z \bar {\cal F}^{\mu \nu}) = 0,
\end{equation}
\begin{equation}
\nabla J^z - J^z(J^z - \bar J^z) - \frac {i}{2}(z - \bar z)
\bar {\cal F} {\cal F}^+ = 0,
\end{equation}
\begin{equation}
R_{\mu \nu} = Tr\{2J^z_{(\mu}\bar J^z_{\nu)} +
i(z - \bar z)({\cal F}_{\mu \lambda}{{{\cal F}^+}_{\nu .}}^{\lambda} +
\bar {\cal F}_{\mu \lambda}{{{\cal F}^T}_{\nu .}}^{\lambda})\},
\end{equation}
where $J^z = \nabla z(z - \bar z)^{-1}$ (see also \cite {stw} for the
EMDA case).

It is not difficult to prove using a straightforward calculation that
these equations are invariant under the transformation
\begin{equation}
z\rightarrow s^T(z^{-1} + l)^{-1}s + r, \qquad
{\cal F}\rightarrow s^{-1}(I + lz){\cal F},
\end{equation}
where the constant matrices $r^T = r$, $l^T = l$ and $s$ are real. The former
formula  from (18) also can be rewritten using the following $2n \times 2n$
matrices
\begin{eqnarray}
m = \left (\begin{array}{crc}
p^{-1} &\quad & p^{-1}q\\
qp^{-1} &\quad & p + qp^{-1}q\\
\end{array}\right )
\end{eqnarray}
(see the similar matrix for the Einstein--Dilaton--Axion theory in
\cite {b})
\begin{eqnarray}
g = \left (\begin{array}{crc}
(s^T)^{-1} & \quad & (s^T)^{-1}r\\
l(s^T)^{-1} & \quad & s + l(s^T)^{-1}r\\
\end{array}\right ),
\end{eqnarray}
as
\begin{equation}
m\rightarrow g^Tmg.
\end{equation}

The above introduced matrix $g$ satisfies the relation $g^Tjg = j$, where
\begin{eqnarray}
j = \left (\begin{array}{crc}
0 & -I\\
I & 0\\
\end{array}\right ),
\end{eqnarray}
and $I$ is the unit matrix,
so that $g \in Sp(2n, R)$ (also from the relations $m^Tjm = j$ and
$m^T = m$ it follows that $m \in Sp(2n, R)/U(n)$, see the third chapter
for details). Thus, the motion equations possess the $Sp(2n, R)$
symmetry for an arbitrary $n \ge 1$.

Not all the symplectic transformations (21) can be represented in the Gauss
decomposition form (20). For example, one can prove that $j \in Sp(2n, R)$,
but only the limit procedure defined by the matrices $l = - r = s \rightarrow
\infty$ allows to obtain the matrix $j$ from Eq. (20). The corresponding map is
$m \rightarrow m^{-1}$; and using complex variables one has:
\begin{equation}
z \rightarrow - z^{-1}, \qquad {\cal F} \rightarrow z{\cal F}.
\end{equation}
It is not difficult to prove that this transformation is actually the
symmetry transformation for the equations (15)--(17).
The map (23) generalizes the strong--weak coupling duality
of the EMDA theory to the case of an arbitrary $n$. It has the same form as
the transformation established for the low energy heterotic string theory
with moduli in four dimensions \cite {s4}.

It is easy to see that only the $GL(n, R)$ subgroup of the complete
$Sp[2(n + 1), R]$ group, defined
by a matrix $s$, preserves the action (12) which can be rewritten as
\begin{equation}
^4S = \int d^4x {\mid g \mid}^{\frac {1}{2}} \{ - R +
Tr [2J^z\bar J^z + i(z{\cal FF}^T - \bar z\bar {\cal F}{\cal F}^+)]\}.
\end{equation}
The situation  is the same one as for the EMDA theory which possesses
the $Sp(2, R) \sim SL(2, R)$ symmetry only for the motion equations
\cite {stw}--\cite {ko}. This symmetry is connected with the notable
$S$--duality for the low energy heterotic string theory.
The subgroup $SL(2, Z)$ of the group $SL(2, R)$ is an exact symmetry for
the heterotic string in four dimensions \cite {s4}.

Also one can see that the discussed models do not possess any analogy with
the $T$--duality transformations \cite {s4}, thus the established
$Sp(2n, R)$ symmetry is the single non--trivial one.
\section{Reduction to Three Dimensions}
In this section we study the stationary case using the classical approach
of Israel and Wilson \cite {iw}. This method also had been explored
for the EMDA theory analysis \cite {ggk}; its application to the string
gravity with moduli fields one can find in \cite {s3}.

First of all, it is convenient to parametrize the four--dimensional line
element as
\begin{equation}
ds_4^2 = g_{\mu \nu}dx^{\mu}dx^{\nu} = f(dt - \omega _mdx^m)^2 -
f^{-1}h_{mn}dx^mdx^n,
\end{equation}
where $m = 1,2,3$.
(In this section all vector variables, as well as the operator $\nabla$,
will be related with the three--metric $h_{mn}$).

Then, the $\mu = m$ component of Eq. (9) allows to introduce the
magnetic potential $u$ according to
\begin{equation}
pF^{mn} + q\tilde F^{mn} = \frac {1}{\sqrt 2}fE^{mnk} u_{,k},
\end{equation}
while for the electric potential $v$ we put
\begin{equation}
F_{m0} = \frac {1}{\sqrt 2} v_{,m}.
\end{equation}
Using the relations \cite {iw} $F^{0m} = \omega _nF^{mn} + h^{mn}F_{n0}$
and $F_{mn} = f^{-2}h_{mk}h_{nl}F^{kl} + 2F_{0[m}\omega _{n]}$, it is possible
to rewrite the $\mu = 0$ component of Eq. (9) as
\begin{equation}
\nabla [f^{-1}(p\nabla v - qp^{-1}(\nabla u - q\nabla v))] =
- f^{-2}\vec \tau \nabla u,
\end{equation}
where $\vec \tau = -f^2\nabla \times \vec \omega$. The second electromagnetic
equation can be obtained from the Bianchi identity $F_{[mn,k]} = 0$; the
result is:
\begin{equation}
\nabla [f^{-1}p^{-1}(\nabla u - q\nabla v)] =
f^{-2}\vec \tau \nabla v.
\end{equation}
The equations (8) and (10) transform to
\begin{eqnarray}
\nabla J^q - J^pJ^q - f^{-1}[(\nabla u - q\nabla v)\nabla v^T +
p\nabla v(\nabla u - q\nabla v)^Tp^{-1}) &=& 0,\\
\nabla J^p + (J^p)^2 + f^{-1}[p\nabla v \nabla v^T -
(\nabla u - q\nabla v)(\nabla u - q\nabla v)^Tp^{-1}] &=& 0.
\end{eqnarray}

Then, the $(_0^m)$--component of the Einstein equations (11) means the
existence of the rotational potential $\chi$
\begin{equation}
\nabla \chi + v^T\nabla u - u^T\nabla v = \vec \tau, \nonumber
\end{equation}
which accordingly with the definition of $\vec \tau$ satisfies
\begin{equation}
\nabla [f^{-2}(\nabla \chi + v^T\nabla u - u^T\nabla v)] = 0;
\end{equation}
while the $(_{00})$--component of Eq. (11) leads to the following equation
for the function $f$:
\begin{eqnarray}
f\nabla ^2f - (\nabla f)^2 + (\nabla \chi + v^T\nabla u - u^T\nabla v)^2
&-&\nonumber\\
f[\nabla v^Tp\nabla v + (\nabla u - q\nabla v)^Tp^{-1}(\nabla u -
q\nabla v)]
&=& 0.
\end{eqnarray}
Finally, the $(^{mn})$--component of Eq. (11) after the lowering of the
three--dimensional indexes takes the form:
\begin{eqnarray}
^3R_{mn} &=& \frac {1}{2} Tr [J^p_mJ^p_n + J^q_mJ^q_n] +
\frac {1}{2}f^{-2}[f_{,m}f_{,n} + \vec \tau _m\vec \tau _n]\nonumber\\
&-& f^{-1}[v^T_{,m}pv_{,n} + (u_{,m} -qv_{,m})^Tp^{-1}(u_{,n}-qv_{,n})],
\end{eqnarray}
where $^3R$ is constructed using the metric $h_{mn}$ and $\vec \tau _m$ is
given by Eq. (32).

One can prove that Eqs. (28)--(31) and (33)--(35) are the
Euler--Lagrange ones for the action
\begin{eqnarray}
^3S =
\int d^3x h^{\frac {1}{2}} \{ &-& ^3R +
\frac {1}{2} Tr[(J^p)^2 + (J^q)^2] + \frac {1}{2}f^{-2}[(\nabla f)^2 +
(\nabla \chi + v^T\nabla u - u^T\nabla v)^2]\nonumber\\ &-&
f^{-1}[\nabla v^Tp\nabla v + (\nabla u - q\nabla v)^T
p^{-1}(\nabla u - q\nabla v)]\}.
\end{eqnarray}
Hence, this system allows the sigma--model representation in the
stationary case.

This result contains the one for the EMDA theory
\cite {ggk}--\cite {gk3}, which admits some compact
matrix formulations. To establish them for an arbitrary $n$, it is
convenient to introduce the complex potentials
\begin{eqnarray}
\Phi &=& u - zv,\\
{\cal E} &=& if - \chi + v^T\Phi,\nonumber
\end{eqnarray}
which together with the matrix $z$, entered before (13), form the set of
Ernst--like
potentials \cite {e1}--\cite {e2} for this problem (see also \cite {gk1} for the
EMDA case). These potentials can be combined into the following
$(n+1)\times (n+1)$ symmetric matrix
\begin{eqnarray}
E = \left (\begin{array}{rcr}
{\cal E}&\qquad &\Phi^T\\
\Phi&\quad &-z\\
\end{array}\right ),
\end{eqnarray}
which provides the K\"{a}hler representation of the model. Namely,
it is easy to check that
\begin{equation}
^3S = \int d^3x h^{\frac {1}{2}}\{ - ^3R + 2Tr(J^E\bar J^E)\},
\end{equation}
where $J^E = \nabla E (E - \bar E)^{-1}$ and the 2--form ${\cal K} =
2iTr({\cal J}^E\wedge \bar {\cal J}^E)$, corresponding to the target space
metric $dl^2 = 2Tr({\cal J}^E\bar {\cal J}^E)$ with ${\cal J}^E =
dE (E - \bar E)^{-1}$, is exact. The K\"ahler formulations of the Einstein
and Einstein--Maxwell theories are well known in the literature
\cite {maz}. The another examples of the K\"ahler systems one can find in
\cite {f}.

The equations following from the action (39)
\begin{equation}
\nabla J^E = J^E(J^E - \bar J^E),
\end{equation}
\begin{equation}
^3R_{mn} = 2Tr(J^E_{(m}\bar J^E_{n)})
\end{equation}
coincide with the Ernst ones \cite {e1} in the case of $n = 0$.
Thus, it is natural to name
the matrix function $E$ as the ``matrix Ernst potential'' for an arbitrary
$n$. This Ernst--like representation for the model provides the above
mentioned complete formal analogy between it and the Einstein theory
in the stationary case.

Then, using the real and imaginary parts of $E = Q + iP$,
\begin{eqnarray}
P = \left (\begin{array}{crc}
f - v^Tpv &\quad & - v^Tp\\
- pv &\quad & - p\\
\end{array}\right ),\qquad
Q=\left (\begin{array}{crc}
v^Tw - \chi &\quad & w^T\\
w &\quad & - q\\
\end{array}\right ),
\end{eqnarray}
where $w = u - qv$, one can rewrite the action (39) as
\begin{equation}
^3S = \int d^3x h^{\frac {1}{2}}\{ - ^3R + \frac {1}{2}Tr[(J^P)^2 +
(J^Q)^2]\},
\end{equation}
with $J^P = (\nabla P)P^{-1}$ and $J^Q = (\nabla Q)P^{-1}$.

Now let us establish an alternative matrix formulation of the problem based
on the use of the original non--target space variables. One can see that
the first equation of motion corresponding to the action (43),
\begin{equation}
\nabla J^Q - J^PJ^Q = 0,
\end{equation}
\begin{equation}
\nabla J^P + (J^Q)^2 = 0,
\end{equation}
\begin{equation}
^3R_{mn} = \frac {1}{2}Tr[J^P_mJ^P_n + J^Q_mJ^Q_n],
\end{equation}
being written as $\nabla [P^{-1}(\nabla Q)P^{-1}] = 0$, ensures the
compatibility condition for the relation
\begin{equation}
\nabla \times \vec \Omega = P^{-1}(\nabla Q)P^{-1}
\end{equation}
which defines the vector matrix $\vec \Omega$. Then, using the relations
(7), (26), (32) and the definition of $\vec \tau$ one obtains the explicit
form of this matrix:
\begin{eqnarray}
\vec \Omega = \left (\begin{array}{ccc}
\vec \omega &\quad & - \sqrt 2(\vec A + A_0\vec \omega)^T\\
- \sqrt 2(\vec A + A_0\vec \omega) &\quad & 2\vec B
+ (\vec A + A_0\vec \omega)A_0^T + A_0(\vec A + A_0\vec \omega)^T\\
\end{array}\right ),
\end{eqnarray}
where $B_m = 2B_{0m}$. It is easy to see that the matrix $\vec \Omega$
consists of the original field variables from Eqs. (1) and (2) which had been
named as ``nondualized'' ones in \cite {ky} for the case of the EMDA theory.

The remaining components of the Kalb--Ramond field
$B_{mn} \equiv \frac {1}{2}E_{mnk}C^k$ satisfy the nondynamical equation
\begin{eqnarray}
&\nabla& \vec C = \{\nabla [\vec B \times \vec \omega ] +
[\vec B - (\vec A + A_0\vec \omega)A^T_0 - A_0(\vec A + A_0\vec
\omega)^T]\nabla \times \vec \omega \nonumber\\
&+&
(\vec A + A_0\vec \omega)\nabla \times (\vec A
+ A_0\vec \omega)^T\} + \nabla \times (\vec A + A_0\vec \omega)(\vec A +
A_0\vec \omega)^T,
\end{eqnarray}
as it follows from Eqs. (7) and the $t$--independance of the $q$.

Then, it is easy to see from Eqs. (47) that the matrices
$P$ and $\vec \Omega$ satisfy the relation
\mbox {$\nabla \times [P\nabla \times (\vec \Omega) P] = 0$}
which also can be rewritten using the matrix current $J^{\vec \Omega} =
P\nabla \times \vec \Omega$ as
\begin{equation}
\nabla \times J^{\vec \Omega} - J^{\vec \Omega}\times J^P = 0.
\end{equation}
This relation together with Eqs. (45) and (46), expressed in terms
of the matrices $P$ and $\vec \Omega$,
\begin{equation}
\nabla J^P + (J^{\vec \Omega})^2 = 0,
\end{equation}
\begin{equation}
^3R_{mn} = \frac {1}{2}Tr[J^P_mJ^P_n + J^{\vec \Omega}_mJ^{\vec \Omega}_n],
\end{equation}
form the set of motion equations for the action
\begin{equation}
^3S = \int d^3xh^{\frac {1}{2}}(- ^3R + \frac {1}{2}Tr[(J^P)^2 -
(J^{\vec \Omega})^2]);
\end{equation}
thus these matrices provide an alternative Lagrange formulation of the
theory under consideration.

To establish the symmetry group of the system, it is convenient to
introduce the
\noindent\\
$2(n + 1)\times 2(n + 1)$ symmetric matrix
\begin{eqnarray}
M = \left (\begin{array}{crc}
P^{-1} &\quad & P^{-1}Q\\
QP^{-1} &\quad & P + QP^{-1}Q\\
\end{array}\right ),
\end{eqnarray}
which possesses the symplectic property
\begin{eqnarray}
M^TJM = J, \qquad {\rm where} \qquad
J = \left (\begin{array}{crc}
0 & -I\\
I & 0\\
\end{array}\right ).
\end{eqnarray}
Then the action (43) obtains the form
\begin{equation}
^3S = \int d^3x h^{\frac {1}{2}}\{ - ^3R + \frac {1}{4}Tr[(J^M)^2]\},
\end{equation}
where $J^M = \nabla M M^{-1}$, and the corresponding equations of
motion are:
\begin{equation}
\nabla J^M = 0,
\end{equation}
\begin{equation}
^3R_{mn} = \frac {1}{4} Tr[J^M_mJ^M_n].
\end{equation}

It is known that in the case of $n = 0$ (for the Einstein theory) a matrix
$M$ belongs to the coset $Sp(2,R)/U(1)$ (or, equivalently, to the
$SL(2, R)/O(2)$) \cite {k1}. Also, it has been established that for the
EMDA theory one has $M\in Sp(4, R)/U(2)$ \cite {gk1}, \cite {gk2}. Hence,
it is natural to suppose that $M\in Sp[2(n + 1), R]/U(n + 1)$ for an
arbitrary $n$.

This supposition really takes place and the local isomorphism can be
established as follows. One can see that a matrix $M$ remains
symplectic and symmetric under the transformation
\begin{equation}
M\rightarrow G^T M G
\end{equation}
with an arbitrary symplectic matrix $G$. It means that the group of the
isometry transformations for the action (56) is isomorphic to
$Sp[2(n + 1), R]$. The set of the corresponding generators $\Gamma = G - I$
can be obtained from (59); the result is:
\begin{eqnarray}
\Gamma = \left (\begin{array}{crc}
-\Gamma _S^T & \Gamma _R\\
\Gamma _L & \Gamma _S\\
\end{array}\right ),
\end{eqnarray}
where $(n + 1)\times (n + 1)$ infinitesimal matrices $\Gamma _R$ and
$\Gamma _L$ are symmetric, and $\Gamma _S$ is an arbitrary matrix of
the same dimension. Then, to obtain the coset, one must remove all the
antisymmetric generators from eq. (60), i. e., to exclude the set of generators
which is defined by the matrix
\begin{eqnarray}
\tilde \Gamma = \left (\begin{array}{crc}
\tilde \Gamma _S & \tilde \Gamma _R\\
- \tilde \Gamma _R & \tilde \Gamma _S\\
\end{array}\right ),
\end{eqnarray}
where $\tilde \Gamma _S^T = - \tilde \Gamma _S$ and $\tilde \Gamma _R^T
= \tilde \Gamma _R$.

From the other hand, the generators $u$ of the group $U(n + 1)$ can be
found from the relation $U^+U = I$ in the infinitesimal case of $U = I + u$.
The result is:
\begin{equation}
u = u_1 + iu_2,
\end{equation}
where $u_1^T = - u_1$ and $u_2^T = u_2$ are the real $(n + 1)\times (n + 1)$
matrices. Now one can prove using a straightforward calculation of the
commutators for the corresponding generators that the waiting isomorphism
can be defined as
\begin{equation}
\tilde \Gamma _S \equiv u_1, \qquad \tilde \Gamma _R \equiv u_2.
\end{equation}

Now let us establish the action of the $Sp[2(n + 1), R]$ group of the
isometry transformations on the Ernst matrix potential $E$. One can see
that the matrix
\begin{eqnarray}
G = \left (\begin{array}{crc}
(S^T)^{-1} & \quad & (S^T)^{-1}R\\
L(S^T)^{-1} & \quad & S + L(S^T)^{-1}R\\
\end{array}\right ),
\end{eqnarray}
with $R^T = R$ and $L^T = L$ possesses the symplectic property and describes
all the belonging to $Sp[2(n + 1), R]$ transformations which can be
continuously connected with the identical one. It is easy to show
that the corresponding transformation for the matrix $E$ is
\begin{equation}
E \rightarrow S^T(E^{-1} + L)^{-1}S + R.
\end{equation}
Thus, the matrix $E$ undergoes the ``matrix--valued $SL(2, R)$''
transformations which are the matrix generalizations of the $SL(2, R)$
ones for the Einstein theory. Namely, the transformation defined by the
matrix
$L$ is the matrix generalization of the Ehlers one \cite {e}, while the
matrices $S$ and $R$ define the transformations which generalize the
rescaling and gauge shift correspondingly for the stationary
Einstein system \cite {k1}.

The transformation (65) has the same form as the transformation for the
matrix $z$ in Eq. (18). It is not difficult to prove that one can identify the
$Sp(2n, R)$ subgroup of the complete group $Sp[2(n + 1), R]$, defined by
the matrices \begin{eqnarray}
L = \left (\begin{array}{crc}
0 & \quad & 0\\
0 & \quad & l
\end{array}\right ),
\qquad
S = \left (\begin{array}{crc}
1 & \quad & 0\\
0 & \quad & s
\end{array}\right ),
\qquad
R = \left (\begin{array}{crc}
0 & \quad & 0\\
0 & \quad & r\\
\end{array}\right ),
\end{eqnarray}
where $l^T = l$ and $r^T = r$,
with the $Sp(2n, R)$ symmetry transformations (18) of the
motion equations (15)--(17) in the stationary case.

As an example of a symplectic transformation which can not be represented
in the form of (64), one can take the matrix $J$. It is easy to see that the
corresponding map is $M \rightarrow M^{-1}$ and its complex representation
has the form $E \rightarrow - E^{-1}$. This transformation can be obtained
from the decomposition formula (64) using the limit procedure as it has been
done in the four--dimensional case. It also provides an analogy  with the
strong--weak coupling duality transformation of the low energy
heterotic string theory with moduli in three dimensions \cite {s3}.

At the end of this section we want to note that all the transformations
discussed here preserve both the actions and the corresponding sets of
motion equations.
\section{Two--Dimensional Case}
In this section we consider some additional properties of the theory
which arise in the stationary and axisymmetric case.

Here one can parametrize the three--dimensional line element in the
Lewis--Papapetrou form:
\begin{equation}
ds_3^2 = e^{2\gamma}(d\rho ^2 + dz^2) + \rho ^2d\varphi ^2.
\end{equation}
Then Eq. (57) transforms to
\begin{equation}
\nabla (\rho J^M) = 0
\end{equation}
which is the Euler--Lagrange equation for the action
\begin{equation}
^2S = \frac {1}{4}\int d\rho dz \rho Tr[(J^M)^2];
\end{equation}
while the Einstein equations (58) become the relations defining the function
$\gamma$:
\begin{eqnarray}
\gamma _{,z} &=& \frac {\rho}{4}Tr[J^M_{\rho}J^M_z],
\\
\gamma _{,\rho} &=& \frac {\rho}{8}Tr[(J^M_{\rho})^2 - (J^M_z)^2]
\nonumber
\end{eqnarray}
(the operator $\nabla$ is connected with the flat metric $\delta _{ab}$ in
this section).

The remarkable fact is that one can obtain the another chiral
matrix for the stationary axisymmetric equations of the problem. To
make it, we firstly consider the formalism based on the use of the
matrices $P$ and $Q$. It is easy to see that the $\varphi$--independent
equations of motion (44)--(45) take the form
\begin{equation}
\nabla (\rho J^Q) - \rho J^PJ^Q = 0,
\end{equation}
\begin{equation}
\nabla (\rho J^P) + \rho (J^Q)^2 = 0,
\end{equation}
and can be derived from the action
\begin{equation}
^2S = \frac {1}{2}\int d\rho dz \rho Tr[(J^P)^2 + (J^Q)^2].
\end{equation}
Also, from Eq. (46) one obtains:
\begin{eqnarray}
\gamma _{,z} &=& \frac {\rho}{2}Tr[J^P_{\rho}J^P_{z} + J^Q_{\rho}J^Q_{z}],
\\
\gamma _{,\rho} &=& \frac {\rho}{4}Tr[(J^P_{\rho})^2 - (J^P_{z})^2 +
(J^Q_{\rho})^2 - (J^Q_{z})^2].
\nonumber
\end{eqnarray}
The equation (71), being written as $\nabla [\rho P^{-1}(\nabla Q)P^{-1}] = 0$,
provides the compatibility condition for the relation
\begin{equation}
\nabla \Omega = \rho P^{-1}(\tilde \nabla Q)P^{-1}
\end{equation}
which defines the symmetric matrix $\Omega$ (here, accordingly with
\cite {k1}, $\tilde \nabla _{\rho } = \nabla _z$ and $\tilde \nabla _z =
- \nabla _{\rho}$). Then, using a straightforward calculation one
obtains that
\begin{equation}
\Omega = \vec \Omega _{\varphi}.
\end{equation}
It is easy to see that the compatibility condition for the relation (75),
rewritten as \mbox {$\nabla Q = - \rho ^{-1}P (\tilde \nabla \Omega) P$},
and the equation (72), expressed in terms of the matrices $P$ and $\Omega$,
\begin{equation}
\nabla (\rho ^{-1}J^{\Omega}) + \rho ^{-1}J^{\Omega}J^P = 0,
\end{equation}
\begin{equation}
\nabla (\rho ^{-1}J^P) + \rho ^{-1}(J^{\Omega})^2 = 0,
\end{equation}
form the Euler--Lagrange set of equations for the action
\begin{equation}
^2S = \frac {1}{2}\int d\rho dz Tr[\rho (J^P)^2 - \rho ^{-1}(J^{\Omega})^2],
\end{equation}
where $J^{\Omega} = P\nabla \Omega$.
Also one can see from Eq. (49), that $\nabla \vec C = 0$ and, fixing a gauge,
it is possible to put $\vec C = 0$ without lose of generality. Then,
for the function $\gamma$ one has:
\begin{eqnarray}
\gamma _{,z} &=& \frac {1}{2}Tr[\rho J^P_{\rho}J^P_{z} -
\rho ^{-1}J^{\Omega}_{\rho}J^{\Omega}_{z}],
\\
\gamma _{,\rho} &=& \frac {1}{4}Tr[\rho ((J^P_{\rho})^2 - (J^P_{z})^2) +
\rho ^{-1}((J^{\Omega}_{\rho})^2 - (J^{\Omega}_{z})^2)].
\nonumber
\end{eqnarray}

The new chiral matrix $N$ can be defined as follows
\begin{eqnarray}
N = \left (\begin{array}{ccc}
P & \quad & - P\Omega\\
- \Omega P & \quad & \Omega P\Omega - \rho ^2P^{-1}\\
\end{array}\right ).
\end{eqnarray}
It satisfies the equation
\begin{equation}
\nabla [\rho J^N] = 0,
\end{equation}
where $J^N = \nabla N N^{-1}$, while the relations for the function $\gamma$
are:
\begin{eqnarray}
\Gamma _{,z} &=& \frac {\rho}{4}Tr[J^N_{\rho}J^N_{z}],\nonumber \\
\Gamma _{,\rho} &=& \frac {\rho}{8}Tr[(J^N_{\rho})^2 - (J^N_{z})^2],
\end{eqnarray}
with
\begin{equation}
\Gamma = \gamma - \frac {1}{2}ln\mid det P\mid + \frac {n + 1}{2}ln\rho.
\end{equation}

It is easy to check that the introduced symmetric matrix $N$ satisfies the
non--group relation
\begin{equation}
NJN = - \rho ^2J
\end{equation}
which is equivalent to the normalization condition $det N = - \rho ^2$ in
the case of the Einstein theory (see also \cite {ky} for the EMDA case).

An analogy between the model under consideration and the Einstein theory
allows to establish one additional property of the system. Namely, the
relation (75) provides the differential correspondence between the
formulations of the problem based on the use of the dualized matrix $Q$
and the original one $\Omega$. The remarkable fact is that also there is
a pure algebraical map
\begin{equation}
P \rightarrow \rho P^{-1}, \qquad Q \rightarrow i\Omega,
\end{equation}
which directly transforms the equations (71)--(72) into the ones (77)--(78).
The corresponding transformation for the function $\gamma$ is:
\begin{equation}
e^{2\gamma ^{\Omega}}
= \frac {\rho ^{\frac {n + 1}{2}}}{\mid det P\mid}e^{2\gamma ^Q},
\end{equation}
where $\gamma ^{\Omega}$ is connected with matrices $P$ and $\Omega$,
and $\gamma ^Q$ is related with the formulation  using $P$ and $Q$.
The map (86) coincides with the Kramer--Neugebauer transformation in the
cases of $n = 0$
\cite {kn} and $n = 1$ \cite {ky} and can naturally preserve this name for an
arbitrary $n$.

Now let us establish the action of the $Sp[2(n + 1), R]$ group on the
matrices $P$ and $\Omega$. To make it one can introduce the matrices
\begin{equation}
E_{\pm} = \Omega \pm \rho P^{-1}.
\end{equation}
It is easy to check that the matrix $N$ remains symmetric and preserves
the property (85) under the transformation $N \rightarrow G^T N G$ with an
arbitrary symplectic matrix $G$. Then, in the case of the matrix $G$,
possessing the decomposition property (64), one has:
\begin{equation}
E_{\pm} \rightarrow S^T(E_{\pm}^{-1} + L)^{-1}S + R.
\end{equation}
Thus, the real nondualized matrix variables $E_{\pm}$ transform under the
$Sp[2(n + 1), R]$ group of transformations as the matrix Ernst potentials
$E$ and $\bar E$.

The entered matrices $E_{\pm}$, being the analogy for the Ernst ones $E$ and
$\bar E$, provide also the compact Lagrange formulation of the problem.
Namely, one can check that the action (79) can be rewritten as
\begin{equation}
^2S = - 2\int d\rho dz \rho Tr[J^{E_+}J^{E_-}],
\end{equation}
where $J^{E_{\pm}} = (E_{\pm} - E_{\mp})^{-1}\nabla E_{\pm} -
\frac {1}{2}\nabla \rho$.

Also one can see that the above established Kramer--Neugebauer map has an
especially simple form in terms of the Ernst matrices
$E, \bar E$ and their real analogies $E_+, E_-$:
\begin{equation}
E\rightarrow iE_+, \qquad \bar E\rightarrow iE_-.
\end{equation}
\section{Conclusion}
In this paper we have considered a class of gravity models with the
Einstein theory as the first representative, the EMDA theory as the second
one, etc. It is shown that all the models
have the same formal properties which are the natural matrix generalizations
of the ones for the Einstein theory in three and two dimensions. Namely, the
substitution $P \rightarrow f, \quad Q \rightarrow \chi$ and $\vec \Omega
\rightarrow \vec \omega$ (or $\Omega \rightarrow \omega$ in the stationary
axisymmetric case) directly transforms all the relations for $n \ge 1$
to the Einstein theory relations. An analogy established here is even a more
complete one, and it is possible to develop the Hauser--Ernst formalism
\cite {he1}, \cite {he3} and to construct the Geroch group \
\cite {g1}--\cite {g2}
in the stationary axisymmetric case for the model with an arbitrary $n$.
\acknowledgments
We would like to thank our colleagues from the Nuclear Physics Institute
for an encouraging relation to our work.

\end{document}